\documentclass[english,10pt,aps,pra,twocolumn,groupedaddress,floatfix]{revtex4-1}
\pdfoutput=1
\usepackage[utf8]{inputenc}
\usepackage[T1]{fontenc}
\usepackage{amsmath}
\usepackage{amssymb}
\usepackage{amsfonts}
\usepackage{bm}
\usepackage{bbm}
\usepackage{epsfig}
\usepackage{grffile}
\usepackage{graphics}
\usepackage[export]{adjustbox}

\usepackage[usenames,dvipsnames]{color}
\definecolor{dblue}{rgb}{0,0.1,.6}

\usepackage{algorithm}
\usepackage{algpseudocode}

\usepackage[colorlinks=true,citecolor=dblue,linkcolor=dblue,urlcolor=dblue]{hyperref}
\usepackage[all]{hypcap}

\newcommand{\id}{\mathbbm{1}}
\newcommand{\Tr}{\operatorname{Tr}}
\newcommand{\bra}{\langle}
\newcommand{\ket}{\rangle}
\newcommand{\mc}[1]{\mathcal{#1}}
\newcommand{\pdag}{{\phantom{\dag}}}
\renewcommand{\vec}[1]{{\boldsymbol{#1}}}
\newcommand{\mri}{\mathrm{i}\mkern1mu}

\newcommand{\M}{\mc{M}}
\newcommand{\N}{\mc{N}}
\newcommand{\E}{\mc{E}}
\newcommand{\A}{\mc{A}}
\renewcommand{\O}{\mc{O}}

\newcommand{\hs}{{\hat{\sigma}}}
\newcommand{\dm}{{\hat{\rho}}}
\newcommand{\hH}{\hat{H}}
\newcommand{\hh}{\hat{h}}
\newcommand{\hL}{\hat{L}}
\newcommand{\hR}{\hat{R}}
\newcommand{\hT}{\hat{T}}
\newcommand{\hU}{\hat{U}}
\newcommand{\hr}{{\hat{r}}}
\newcommand{\hu}{{\hat{u}}}
\newcommand{\hw}{{\hat{w}}}
\newcommand{\hA}{\hat{A}}
\newcommand{\hB}{\hat{B}}
\newcommand{\hd}{\hat{d}}
\newcommand{\hg}{\hat{g}}
\newcommand{\hp}{\hat{p}}
\newcommand{\heta}{\hat{\eta}}

\newcommand{\vg}{\vec{g}}
\newcommand{\vp}{\vec{p}}
\renewcommand{\vr}{\vec{r}}
\newcommand{\vs}{\vec{s}}
\newcommand{\vu}{\vec{u}}
\newcommand{\vy}{\vec{y}}
\newcommand{\RR}{\mathbb{R}}
\newcommand{\CC}{\mathbb{C}}
\newcommand{\U}{\operatorname{U}}
\renewcommand{\Re}{\operatorname{Re}}
\newcommand{\veps}{\varepsilon}

\newcommand{\duke} {Department of Physics, Duke University, Durham, North Carolina 27708, USA}
\newcommand{\dqc}  {Duke Quantum Center, Duke University, Durham, North Carolina 27701, USA}

\newcommand{\Title} {Quantum-classical eigensolver using multiscale entanglement renormalization}
\newcommand{\Authors}
{
\author{Qiang Miao}
\affiliation{\duke}
\affiliation{\dqc}
\author{Thomas Barthel}
\affiliation{\duke}
\affiliation{\dqc}
}
\newcommand{\Date} {July 22, 2021}

\begin{document}

\title{\Title}
\Authors

\begin{abstract}
We propose a variational quantum eigensolver (VQE) for the simulation of strongly-correlated quantum matter based on a multi-scale entanglement renormalization ansatz (MERA) and gradient-based optimization. This MERA quantum eigensolver can have substantially lower computation costs than corresponding classical algorithms. Due to its narrow causal cone, the algorithm can be implemented on noisy intermediate-scale quantum (NISQ) devices and still describe large systems. It is particularly attractive for ion-trap devices with ion-shuttling capabilities. The number of required qubits is system-size independent, and increases only to a logarithmic scaling when using quantum amplitude estimation to speed up gradient evaluations. Translation invariance can be used to make computation costs square-logarithmic in the system size and describe the thermodynamic limit. We demonstrate the approach numerically for a MERA with Trotterized disentanglers and isometries. With a few Trotter steps, one recovers the accuracy of the full MERA.
\end{abstract}

\date{\Date}
\maketitle

\section{Introduction}
The complexity of quantum many-body systems makes it a formidable challenge to understand the properties of quantum matter, in particular in strongly correlated regimes where perturbative approaches fail.
Hence, powerful classical simulation techniques like quantum Monte Carlo \cite{Foulkes2001-73,Suzuki1977-58,Syljuasen2002-66,Prokofev1998-81} and tensor networks states (TNS) \cite{Baxter1968-9,White1992-11,Niggemann1997-104,Verstraete2004-7,Vidal-2005-12,Orus2014-349} have been developed. A strength of TNS techniques is that they are also applicable for frustrated quantum magnets and fermionic systems \cite{Barthel2009-80,Corboz2009-80,Kraus2009_04,Corboz2009_04,Pineda2009_05}, where quantum Monte Carlo is hampered by the negative-sign problem \cite{Loh1990-41,Troyer2005}. These classes of systems include candidate spin liquid materials \cite{Balents2010-464,Zhou2017-89,Shimizu2003-91,Pratt2011-471,Banerjee2016-15}, fractional quantum Hall physics \cite{Stormer1999-71,de-Picciotto1997-389}, and high-temperature superconductors \cite{Bednorz1986-64,Leggett2006-2}.

Consider a lattice system with $N$ sites, each associated with a site Hilbert space of dimension $d$ such that the total Hilbert space has dimension $d^N$. The idea of TNS is to approximate the many-body state by a network of partially contracted tensors. The tensors may carry physical indices that label site basis states and additional bond indices of dimension $\chi$ which are contracted with corresponding indices of other tensors. The structure of the network and the required bond dimension $\chi$ are adapted to the entanglement structure in the system. Typically, the more entangled a system is, the larger $\chi$ needs to be in order to achieve a desired approximation accuracy. To approximate the ground state of a given model $\hH$ by a TNS $|\Psi\ket$, one minimizes the energy $\bra\Psi|\hH|\Psi\ket/\|\Psi\|^2$ with respect to the tensor elements. The beauty of the approach is that computation costs for optimization steps are reduced from exponential in $N$ to polynomial in $N$. In particular, they are linear in $N$ for matrix product states (MPS) \cite{Baxter1968-9,Fannes1992-144,White1992-11,Rommer1997,Schollwoeck2011-326}, projected entangled pair states (PEPS) \cite{Niggemann1997-104,Nishino2000-575,Martin-Delgado2001-64,Verstraete2004-7,Verstraete2006-96}, and the multi-scale entanglement renormalization ansatz (MERA) \cite{Vidal-2005-12,Vidal2006}. For homogeneous MERA, one can reduce the cost to $\O(\log N)$ and even access the thermodynamic limit $N\to\infty$. However, the classical computation time may scale with a high power of the bond dimension $\chi$. While it is only $\O(\chi^3)$ for MPS in one-dimensional (1D) systems \cite{White1992-11,Rommer1997}, it is $\O(\chi^{7\dots 9})$ for 1D MERA \cite{Evenbly2013}, $\O(\chi^{10\dots 12})$ for 2D PEPS \cite{Jordan2008-101,Orus2009_05}, and $\O(\chi^{16\dots 28})$ for 2D MERA \cite{Cincio2008-100,Evenbly2009-102}. Hence, practicable $\chi$ are usually rather small, which limits the approximation accuracy.

In this work, we propose and analyze a hybrid quantum-classical variational eigensolver \cite{McClean2016-18} to overcome these limitations, where many-body ground states are approximated by adapted MERA states and (small) quantum computers are employed to efficiently execute tensor contractions. In this context, MERA have four advantages over other TNS: (i) MERA can be applied for systems with any number of spatial dimensions, (ii) all tensors are unitary or isometric, which allows for a rather direct implementation on quantum computers, (iii) MERA expectation values for local operators depend only on narrow causal cones such that they can be evaluated exactly and large systems can be simulated on noisy intermediate-scale quantum (NISQ) devices, and (iv) sets of MERA are closed which implies that optimizers always exist \cite{Barthel2022-112}. Also, MERA optimizations are not hampered by barren plateaus \cite{Barthel2023_03,Miao2023_04}.

Quantum algorithms using MPS for 1D systems were recently suggested in Refs.~\cite{Barratt2021-7,Liu2019-1,FossFeig2021-3,Smith2022-4,Chertkov2022-18}.
Two prominent quantum-computing platforms are superconducting qubits \cite{Schoelkopf2008-451,Devoret2013-339} and ions in electromagnetic traps \cite{Cirac1995-74,Blatt2008-453}. Ion-trap systems with qubit-shuttling capabilities \cite{Rowe2002-2,Hensinger2006-88,Walther2012-109}
are particularly interesting for the quantum MERA scheme.
\begin{figure*}[t]
	\includegraphics[width=1\textwidth]{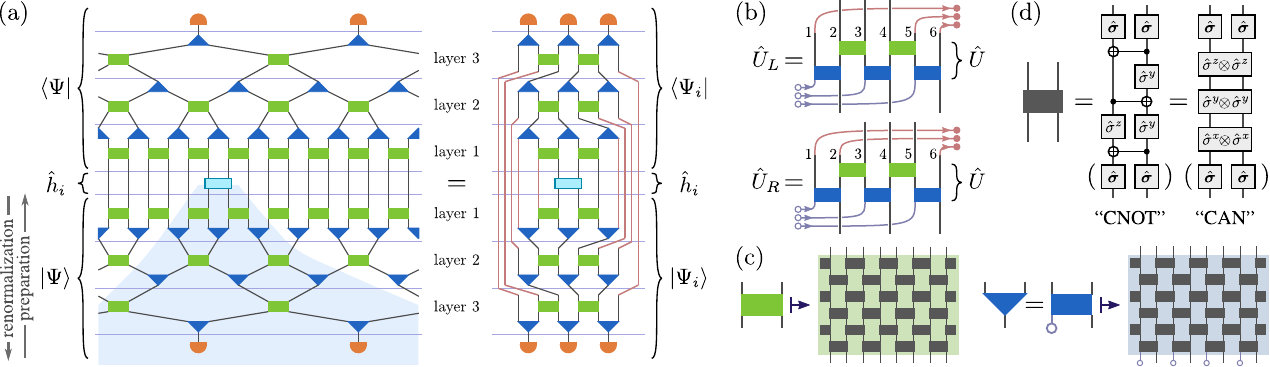}
	\caption{\label{fig:TMERA}\textbf{TMERA structure and implementation.} (a) Expectation value $\bra\Psi|\hh_i|\Psi\ket$ for a homogeneous binary 1D MERA $|\Psi\ket$ with $T=3$ layers. The MERA consists of disentanglers (boxes) and isometries (triangles). Contraction lines between the tensors correspond to renormalized site vector spaces with dimension $\chi$. The shaded region indicates the causal cone for a two-site operator $\hh_i$. The cone has width three, i.e., contains at most three renormalized sites in each layer. Only causal-cone states like $|\Psi_i\ket$, associated with $\hh_i$, need to be generated on the quantum computer.
	(b) Isometries can be realized as unitaries where some input qubits are initialized in a reference state like $|0\ket$ (open circles). Such $|0\ket$ qubits can be moved in and others (filled circles) can be moved out of the quantum register or reset after applying the layer-transition maps.
	(c) A Trotter structure with $t$ steps is imposed on each MERA tensor, making it a circuit of two-qubit gates. Here, $t=3$ and $\chi=16$, i.e., $q=4$ qubits per contraction line.
	(d) Each Trotter gate can be implemented using CNOTs and single-qubit rotations or, equivalently, single and two-qubit rotations. The specified Pauli operators generate the corresponding rotations \eqref{eq:rotation}, and $\hat{\vec{\sigma}}$ refers to a general single-qubit rotation.}
\end{figure*}

\section{MERA adapted for quantum computers}
A MERA \cite{Vidal-2005-12,Vidal2006} is a hierarchical TNS motivated by the real-space renormalization group \cite{Kadanoff1966-2,Jullien1977-38,Drell1977-16}: In each renormalization step $\tau=1,\dotsc, T$, unitaries with small spatial support are applied to disentangle the system to some extent, before isometries are applied in order to map a block of $b$ sites into a new renormalized site. In the process, states that are not important for the representation of the ground state are discarded. With a branching ratio of $b$, this process ends with one or a few renormalized sites after $T\sim\log_b N$ steps, and the resulting few-site problem can be solved exactly. While the physical site Hilbert spaces have dimension $d$, Hilbert spaces of renormalized sites have dimension $\chi$. Seen in reverse, the renormalization group scheme defines a many-body state $|\Psi\ket$. This state is a MERA with bond dimension $\chi$. It consists of $T$ layers, each comprising the unitary disentanglers and isometries of a renormalization step. Optimizing the tensor elements to minimize the energy expectation value $E=\bra\Psi|\hH|\Psi\ket$, one obtains a groundstate approximation.

In principle, it is straightforward to prepare a MERA $|\Psi\ket$ on quantum computers. Assume $\chi = 2^q$ such that every renormalized site corresponds to $q$ qubits. A disentangler that acts on $n$ (renormalized) sites is a $\chi^n\times \chi^n$ unitary acting on $nq$ qubits. It can be decomposed into a circuit of $\O(4^{nq})$ single-qubit and CNOT gates \cite{Barenco1995-52,Moettoenen2004-93,Shende2006-25}. An isometry that maps $n$ sites into $m>n$ can be implemented as a unitary acting on $nq$ qubits and $(m-n)q$ additional ones initialized in state $|0\ket$. It requires $\O(2^{(n+m)q})$ single-qubit and CNOT gates \cite{Itan2016-93}.

For simplicity, we assume a Hamiltonian $\hH=\sum \hh_i$ with finite-range interaction terms $\hh_i$. As exemplified in Fig.~\ref{fig:TMERA}a, many tensors cancel in expectation values $\bra\Psi|\hh_i|\Psi\ket$ due to their isometric property. The causal cone of $\hh_i$ comprises all tensors that can influence the expectation value, and we define $|\Psi_i\ket$ as the corresponding causal-cone TNS such that $\bra\Psi|\hh_i|\Psi\ket=\bra\Psi_i|\hh_i|\Psi_i\ket$. For a binary 1D MERA, disentanglers act on $n=2$ sites and the cost to evaluate $\bra\Psi_i|\hh_i|\Psi_i\ket$ would scale in $q$ as $\O(4^{2q})$. Hence, the cost for the evaluation of an energy gradient would scale as $\O(4^{4q}=\chi^{8})$. This is only a modest improvement over the scaling  $\O(\chi^9)$ of the classical computation time. As discussed in Appx.~\ref{sec:Complexity}, the differences are generally more pronounced in higher dimensions. For example, the quantum and classical gradient evaluation costs for the 2D $2\times 2\mapsto 1$ MERA of Ref.~\cite{Cincio2008-100} scale as $\O(4^{8q}=\chi^{16})$ and $\O(\chi^{28})$, respectively. In any event, one also needs to account for the required number of measurement samples in the quantum case, and the quantum computational complexity can be reduced drastically by imposing further structure on the MERA.

\section{Trotterized tensors}\label{sec:Trotterization}
There are many options for substructures. Here, we choose to impose a Trotter structure on the MERA tensors. In particular, they shall consist of $t$ Trotter steps, each comprising local unitary gates that act on, say, two nearest-neighbor qubits; see Fig.~\ref{fig:TMERA}c. For tensors that act on $n$ (renormalized) sites, each Trotter step consists of $\O(nq)$ local gates. For a Trotterized MERA (TMERA) with $T$ layers, the measurement of a local expectation value $\bra\Psi_i|\hh_i|\Psi_i\ket$ then requires $\O(Tt)$ time on the quantum computer. We will see that, using translation invariance, the measurement of the energy gradient requires $\O(T^2t^2nq)$ time.
As we will also see in benchmark simulations, the local unitary gates approach identities when increasing the number $t$ of Trotter steps. This establishes a connection to Trotterization as used in time evolution problems \cite{Trotter1959,Suzuki1976-51,Barthel2020-418,Childs2021-11}.

\section{Hybrid optimization algorithm}
In classical computations, MERA states are optimized by evaluating the so-called environment for each tensor and updating tensors one by one \cite{Evenbly2009-79}. On a quantum computer, we can only measure observables, and the tensor environment is not accessible. The hybrid algorithm works as follows. The Trotter gates can be written as small circuits, parametrized through the angles $\vec{\theta}=(\theta_1,\theta_2\dotsc)$ of rotations 
\begin{equation}\label{eq:rotation}\textstyle
 	\hR_\hs(\theta):=e^{-\mri\theta \hs/2}=\id\cos\frac{\theta}{2}-\mri\hs\sin\frac{\theta}{2}
\end{equation}
with respect to Hermitian unitary operators $\hs$ like the Pauli matrices $\{\id,\hs^x,\hs^y,\hs^z\}$ or tensor products thereof. A standard choice is depicted in Fig.~\ref{fig:TMERA}d. It comprises three CNOT gates, one $\hs^z$ and two $\hs^y$ single-qubit rotations, as well as four general single-qubit gates \cite{Vatan2004-69,Shende2004-69}. The number of angles per Trotter gate agrees with $\dim\operatorname{SU}(4)=15$ and reduces to nine angles when exploiting the unitary gauge freedoms in the TMERA.
The energy gradient $\partial_{\vec{\theta}}E$ can be evaluated by measuring
\begin{equation}\label{eq:grad}\textstyle
	\partial_{\theta_j} E =  \frac{1}{2} \big[ E(\theta_j+\pi/2)- E(\theta_j- \pi/2)\big],
\end{equation}
where all angles except for $\theta_j$ are kept fixed \cite{Li2017-118,Guerreschi2017_01,Mitarai2018-98}.
A derivation is given in Appx.~\ref{sec:Gradients}.
The energy can now be minimized by a gradient-based algorithm like L-BFGS \cite{Nocedal2006,Liu1989-45}.

In experiments, two-qubit gates are typically much more costly than single-qubit gates. 
For ion-trap and superconducting systems, Refs.~\cite{Bruzewicz2019-6,Sheldon2016-93,McKay2019-122,Kjaergaard2020-11} specify typical single-qubit gate times of $\sim 10$ $\mu$s
and $\sim 30$ ns,
respectively, whereas two-qubit gates require $\sim 100$ $\mu$s
and $\sim 200$ ns,
respectively.
The CNOT parametrization (Fig.~\ref{fig:TMERA}d) of the Trotter gates has the drawback that CNOT gates require two-qubit rotations with large angles. In the ion-trap and superconducting systems, CNOT is implemented using an effective Ising $\hs^\alpha\otimes\hs^\alpha$ interaction with rotation angle $\theta=\pi/2$ \cite{Soerensen1999-82,Debnath2016-536,Maslov2017-19,Kjaergaard2020-11}.
A better choice is then the canonical (CAN) parametrization in Fig.~\ref{fig:TMERA}d that comprises three native $\hs^\alpha\otimes\hs^\alpha$ rotations ($\alpha=x,y,z$) and four general single-qubit rotations \cite{Kraus2001-63,Zhang2003-67}. The benchmark simulations, discussed below, show that the occurring two-qubit angles for this parametrization are rather small.
Furthermore, we find that the optimization actually works best in a parametrization-free fashion. Such a Riemannian quasi-Newton method on quantum circuits is described in Appx.~\ref{sec:Riemannian}.

\section{Translation invariance}\label{sec:TranslationInv}
For translation-invariant systems, the interaction terms $\hh_i$ are translates of the same operator $\hh$. Correspondingly, we can reduce the number of variational parameters. A homogeneous MERA has translation-invariant layers, i.e., each layer consists of repeating identical groups of tensors.
A binary 1D MERA, for example, is then characterized by a single disentangler and a single isometry for each layer. For a heterogeneous system, the derivatives \eqref{eq:grad} can be evaluated by measuring expectation values for all terms $h_i$ that have the tensor of angle $\theta_j$ in their causal cone. In total, this requires $\O(NT=N\log_b N)$ measurements. With translation invariance, this can be reduced to $\O(T)$ by either using classical random bits or introducing auxiliary qubits:
For the 1D case illustrated in Fig.~\ref{fig:TMERA}, there are two unitary transition maps $\hU_L$ and $\hU_R$. Either of them has to be applied to progress in the preparation of the causal-cone state $|\Psi_i\ket$ from layer $\tau$ to $\tau-1$. The specific sequence depends on the location $i$ of the interaction term. In order to evaluate the energy density $e:=\frac{1}{N}\bra\Psi|\hH|\Psi\ket=\frac{1}{N}\sum_i\bra\Psi_i|\hh_i|\Psi_i\ket$ for the entire system at once, we can replace $\hU_{L,R}$ by their convex combination such that we obtain the state on layer $\tau-1$ as
\begin{equation}\label{eq:decending}\textstyle
	\dm^{(\tau-1)}=\frac{1}{2}\left(\hU_L^\pdag \dm^{(\tau)} \hU_L^\dag + \hU_R^\pdag \dm^{(\tau)} \hU_R^\dag\right).
\end{equation}
For experiments, such quantum channels can be implemented by randomly selecting $\hU_L$ or $\hU_R$ in each transition. Practically, this constant reprogramming of the hardware in gradient evaluations can be slow. So, alternatively, the channels can be lifted to fixed unitary evolutions on a larger Hilbert space \cite{Stinespring1955-6,Nielsen2000}. For Eq.~\eqref{eq:decending}, adding a single auxiliary qubit per layer is sufficient. Initializing it in the state $(|0\ket+|1\ket)/\sqrt{2}$ and applying $\hU_L$ or $\hU_R$ conditioned on the auxiliary qubit realizes the channel \eqref{eq:decending}. Appendix~\ref{sec:homogMERA} gives details on the efficient realization of layer-transition maps for various MERA.

Homogeneous MERA are necessarily defined with periodic boundary conditions. Let the linear system sizes $L_x,L_y\dotsc$ be large enough such that the causal cone of any local interaction term $\hh_i$ does not close upon itself along any of the spatial dimensions. Then, repeating the MERA tensor network in any spatial direction ($L_\alpha\mapsto nL_\alpha$) with accordingly adapted boundary conditions defines families of homogeneous MERA, which all have the same energy density $e$. In particular, the results capture the thermodynamic limit $N\to\infty$.

\section{Qubit resets and ion shuttling}
\begin{figure*}[t]
	\centering
	{\small (a)}\includegraphics[width=0.44\textwidth,valign=t]{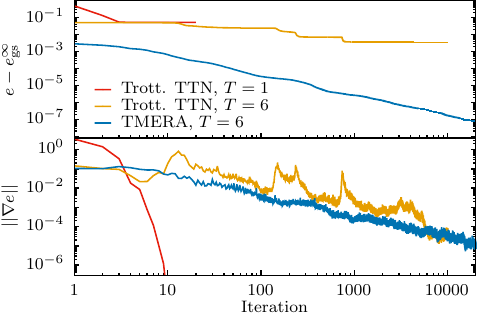}\hspace{0.03\textwidth}
	{\small (b)}\includegraphics[width=0.44\textwidth,valign=t]{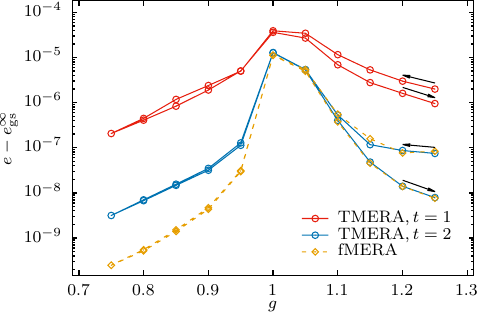}
	\caption{\label{fig:tIsing}\textbf{Benchmark simulations.} The plots show the convergence and accuracy of energy densities $e$ for the transverse Ising model \eqref{eq:tIsing} using homogeneous modified binary MERA with $T=6$ layers and bond dimension $\chi=8$ ($q=3$).
	Panel (a) shows the convergence at $g = 1.25$ for TMERA with $t=2$, starting from the product state with state $e^{-\mri\frac{\pi}{8}\hs^z}e^{-\mri\frac{\pi}{8}\hs^y}|\uparrow{}\ket$ on every site. In first optimization phases, disentanglers are removed (set to $\id$), i.e., Trotterized tree tensor networks (TTN) \cite{Shi2006-74,Murg2010-82} are optimized. The resulting state is used to initialize the TMERA optimization.
	(b) Energy accuracy of optimized TMERA and MERA with full tensors (fMERA) as a function of the field strength $g$. Local minima are avoided by scanning from $g=1.25$ to $g=0.75$ and back. 
	Especially in the paramagnetic phase ($g>1$) and at the critical point $g=1$, the accuracy of the fMERA is recovered with only $t=2$ Trotter steps per tensor.}
\end{figure*}
An attractive feature of the proposed quantum-classical TMERA algorithm is that, while we can simulate large systems, at any stage, only a system-size-independent number of qubits need to be acted upon with the unitary gates. When evaluating observables or gradients, as we progress from layer to layer, only the qubits inside the causal cone need to be in the quantum register. These are, e.g., $4q$ qubits for 1D binary and ternary MERA, and $14q$ qubits for the 2D $2\times 2\mapsto 1$ MERA of Ref.~\cite{Cincio2008-100}.
In every layer transition, some contraction lines (groups of $q$ qubits) leave the causal cone. The same number of (new) qubits, initialized in state $|0\ket$, are needed to realize the isometries of the next MERA layer.
When space efficiency is the highest priority, one can reset the qubits \cite{Reed2010-96,Magnard2018-121,Egger2018-10,Schindler2011-332,Gaebler2021-104} that exit the causal cone to $|0\ket$ for reuse. Auxiliary qubits can be reset as well.
When one wants to minimize execution times, one can employ quantum amplitude estimation (QAE) \cite{Knill2007-75,Wang2019-122} in the gradient evaluations. For a preparation $|\Psi_i\ket=\hU_i|0,\dotsc,0\ket$ of the causal-cone state, QAE requires application of powers $(\hU_i)^m$. In this case, one cannot employ the (non-unitary) mid-circuit qubit resets. While the total number of required qubits is then logarithmic in the system size, still, only causal-cone qubits need to reside in the quantum register, and the others can be moved to a quantum memory. In ion-trap systems, this can be accomplished by shuttling as demonstrated in Refs.~\cite{Rowe2002-2,Hensinger2006-88,Walther2012-109}.
\begin{figure}[t]
	\includegraphics[width=0.95\columnwidth]{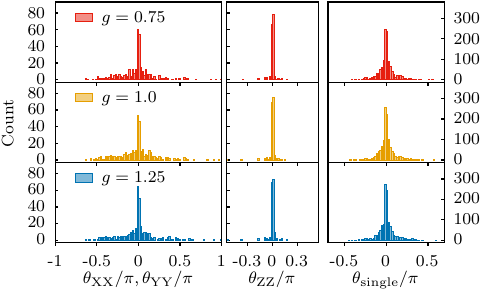}
	\caption{\label{fig:angles}\textbf{Angle distributions in converged TMERA.} Trotter gates can be parametrized in the canonical (CAN) form based on four single-qubit rotations and three Ising-interaction gates (XX, YY, and ZZ). All angles are found to be peaked around 0, which is favorable for the experimental realization. The plots show angle distributions for converged TMERA from Fig.~\ref{fig:tIsing} with $t=2$ Trotter steps.}
\end{figure}

\section{Benchmark simulations and scanning}
To demonstrate and benchmark TMERA, we simulate the 1D transverse-field Ising model
\begin{equation}\label{eq:tIsing}\textstyle
	\hH = -\sum_i \hat{\sigma}^x_i\hat{\sigma}^x_{i+1} + g\sum_i\hat{\sigma}^z_i.
\end{equation}
It has a critical point at $g=1$ with the paramagnetic phase for $g>1$ and the ferromagnetic phase for $g<1$.
Figure~\ref{fig:tIsing} shows results for homogeneous TMERA with the modified binary network structure \cite{Evenbly2013}, using an L-BFGS optimization. The TMERA energy densities $e$ are compared to the exact infinite-system value $e^\infty_\text{gs}$. The left panel shows the convergence for $g=1.25$, which quickly reaches a high accuracy. The right panel, shows TMERA accuracies for $0.75\leq g\leq 1.25$. Local minima are avoided through \emph{scanning}, i.e., starting at $g=1.25$, $g$ is lowered in steps, and the converged TMERA of the previous step is used to initialize the optimization of the next. Upon reaching $g=0.75$, we start scanning back to $g=1.25$. The numerical results confirm that a few Trotter steps $t$ are sufficient to reach accuracies comparable to the full (non-Trotterized) MERA. In particular, $t=2$ gives already excellent results for $\chi=2^3$.

For the experimental implementation, the Trotter gates can be expressed in the CAN representation. Figure~\ref{fig:angles} shows distributions of the rotation angles in the converged TMERA at different $g$. They are peaked at small angles.
This remains true even for the critical point $g=1$. The fact that most angles are small implies that these quantum gates can be executed quickly or at correspondingly higher fidelity.

\section{Computation cost and accuracy}\label{sec:CostAccuracy}
Exploiting translation invariance, the $\O(Ttq)$ components \eqref{eq:grad} of the energy gradient can be evaluated by preparing the corresponding causal-cone states, which costs $\O(Tt)$ time, and then projectively measuring the local interaction term $\hh$. With $N_s$ samples per term, the statistical error of the gradient and, hence, the achievable energy accuracy scale as $\epsilon\propto 1/\sqrt{N_s}$. Thus, the quantum cost for each TMERA optimization step is $\O(T^2t^2 q/\epsilon^2)$. Using QAE \cite{Knill2007-75,Wang2019-122}, the cost reduces to $\O(T^2t^2 q \log(1/\epsilon) /\epsilon)$ while increasing the circuit depth by a factor $\O(1/\epsilon)$.
Our simulations show that the error $\epsilon$ decreases according to a power law $\epsilon\sim (t^2 q)^{-\alpha}$. For fixed $q$, $\epsilon$ decreases until reaching the accuracy of the MERA with full tensors (fMERA) of bond dimension $\chi=2^q$. Upon approaching the saturation, one should increase $q$. The fMERA computation cost also follows a power law $\O(T\chi^r)=\O(\epsilon^{-r/\beta})$, where exponent $r$ is determined by the contraction cost and exponent $\beta$ by the relation between $\chi$ and accuracy $\epsilon$. For 1D MERA, model-dependent exponents $\beta\approx 3.8\dots 6.8$ have been reported \cite{Evenbly2013}. In simulations of the critical bilinear-biquadratic spin-1 chain \cite{Uimin1970-12,Lai1974-15,Sutherland1975-12,Laeuchli2006-74,Binder2020-102} with modified binary MERA ($r=7$), we find that the QAE cost $\O(\epsilon^{-1-1/\alpha})$ is already lower than the classical fMERA cost $\O(\epsilon^{-r/\beta})$, providing a polynomial advantage; see Appx.~\ref{sec:Complexity-BLBQ1}. As the exponent $r$ for the classical fMERA cost is very large for higher-dimensional systems ($r\geq 16$), the quantum algorithm should substantially outperform the classical simulations for models in $\geq 2$ spatial dimensions. It is numerically very expensive to determine the scaling exponents for higher-dimensional systems and further investigations on this subject are needed.

\section{Discussion}
The presented TMERA quantum eigensolver allows for the approximation of many-body ground states with a system-size independent number of qubits, and it can substantially outperform classical MERA simulations.
The appendices provide details on the computational complexity for different MERA network structures (Appx.~\ref{sec:Complexity}), the realization of layer-transition maps for homogeneous TMERA  (Appx.~\ref{sec:homogMERA}), optimization methods (Appx.~\ref{sec:Gradients}) including a Riemannian version of the L-BFGS algorithm, and the influence of different Trotter gate parametrizations (Appx.~\ref{sec:localMinimia}).

Reference~\cite{Kim2017_11} discusses DMERA, which are a special type of TMERA, where the number of Trotter steps $D$ in each layer is directly linked to the width $\sim 2D$ of the causal cone. The TMERA that we consider here have more structure, which allows one to tune these quantities independently, and the imposed structure admits a more direct comparison with the typical MERA used in classical simulations. The optimization based on the simultaneous perturbation stochastic approximation (SPSA), suggested in Ref.~\cite{Kim2017_11}, is considerably less efficient because the energy derivative is evaluated only along random directions in the high-dimensional search space. Our gradient-based approach and the described utilization of translation invariance can be applied for any TMERA, including DMERA. Conversely, the robustness to noise as analyzed in Ref.~\cite{Kim2017_11} also applies generally to TMERA.

The decomposition of the MERA tensors into layers of nearest-neighbor Trotter gates is natural but not necessary. In future research, one could explore other network topologies to leverage, e.g., the all-to-all connectivity of ion-trap systems \cite{Wright2019-10,Linke2017-114} and to increase the expressiveness of TMERA at fixed cost. One could also consider other gate types, especially those that are naturally available in prominent quantum-computing architectures. An example are multi-qubit M\o{}lmer-S\o{}rensen gates \cite{Moelmer1999-82,Soerensen2000-62}.
For an implementation on present-day devices, small two-qubit rotation angles are desirable. Hence, it will be interesting to explore how the angles and the TMERA accuracy are affected by adding large-angle penalty terms to the energy functional. See the follow-up paper Ref.~\cite{Miao2023_03} for further analysis of TMERA.

\emph{Note added.} -- During the long review of this paper, Refs.~\cite{Haghshenas2022-12} and \cite{Haghshenas2023_05} appeared. The first studies the expressiveness of TMERA numerically; the second is an experimental demonstration, measuring critical correlations in preoptimized TMERA with bond dimension $\chi=4$ using an ion-trap system.

\begin{acknowledgments}\vspace{-0.5em}
We gratefully acknowledge helpful discussions with 
Marko Cetina, Kenneth R.\ Brown, Christopher R.\ Monroe, Jungsang Kim, Iman Marvian, Sarah Brandsen, and Yikang Zhang,
and support through US Department of Energy grant DE-SC0019449.
\end{acknowledgments}

\appendix

\renewcommand{\thesection}{\Alph{section}}
\renewcommand{\thesubsection}{\thesection.\arabic{subsection}}
\makeatletter
\renewcommand{\p@subsection}{}
\renewcommand{\p@subsubsection}{}
\makeatother

\section{Computational complexity for 1D and 2D MERA}\label{sec:Complexity}
\begin{figure*}[t]
	\includegraphics[width=0.99\textwidth]{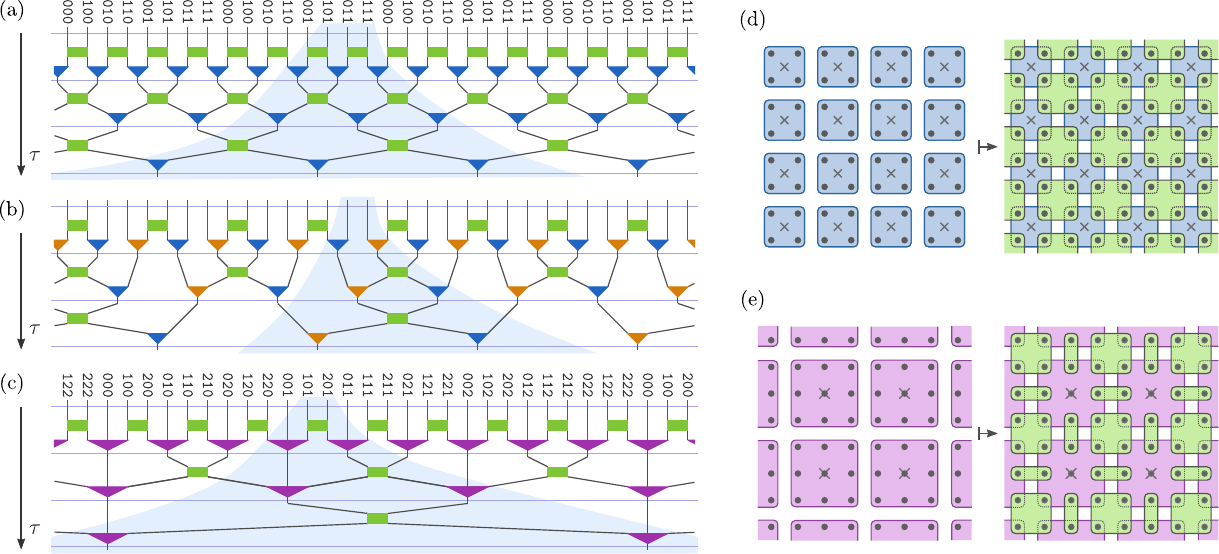}
	\caption{\label{fig:MERAs}\textbf{1D and 2D MERA tensor networks.} Panels (a-c) each show three layers of a 1D MERA, where the preparation direction (decreasing $\tau$) is upwards and causal cones for local operators are indicated: (a) the 1D binary MERA, (b) a  modified 1D binary MERA \cite{Evenbly2013}, and (c) the 1D ternary MERA. Panels (d) and (e) each show one layer of a 2D MERA, where we progress in the preparation direction (decreasing $\tau$) from left to right. (d) For the shown 2D $2\times 2\mapsto 1$ MERA \cite{Cincio2008-100}, isometries map renormalized sites (crosses) into blocks of $2\times 2$ sites (dots) before disentanglers are applied to shifted $2\times 2$ site blocks. (e) For the shown 2D $3\times 3\mapsto 1$ MERA \cite{Evenbly2009-79}, isometries map a renormalized sites (crosses) into blocks of $3\times 3$ sites (dots) before one applies $4$-site and two $2$-site disentanglers.}
\end{figure*}
Let us discuss the quantum computational complexity for different TMERA in one and two dimensions and compare to the corresponding classical simulation costs. Figure~\ref{fig:MERAs} shows the considered MERA networks: the 1D binary MERA, a  modified 1D binary MERA \cite{Evenbly2013}, the 1D ternary MERA, a 2D $2\times 2\mapsto 1$ MERA \cite{Cincio2008-100}, and a 2D $3\times 3\mapsto 1$ MERA \cite{Evenbly2009-79}.
We use the following labels:
\begin{itemize}
\itemsep0em
 \item $b$ denotes the MERA branching ratio.
 \item $T$ denotes the number of layers.
 \item $\tau=1,\dots,T$ labels layers with layer $\tau=1$ acting on the physical sites, i.e., $\tau$ increases in the renormalization direction.
 \item $\chi=2^q$ denotes the bond dimension with $q$ being the corresponding number of qubits per renormalized site.
 \item $A$ denotes the cross section of the causal cone for local operators, defined as the maximum number of renormalized sites inside the causal cone at any layer interface ($\tau\to\tau\pm 1$).
 \item $t$ denotes the number of Trotter steps for each tensor in TMERA (or an upper bound).
\end{itemize}
For heterogeneous MERA, the total number of sites is denoted by $N$ and assumed to be $\sim b^T$.

\subsection{Classical time complexity}\label{sec:ComplexityClassic}
The costs for optimizing MERA on classical computers are determined by the cost of computing the so-called environment of a tensor. This is in turn determined by the cost of applying a renormalization step ($\tau\mapsto \tau+1$) to a local interaction term or, equivalently, for propagating a reduced density matrix inside the causal cone in the preparation direction ($\tau\mapsto \tau-1$). In each step, one needs to contract the tensors of disentanglers and isometries and trace out sites that leave the causal cone. The costs for these operations, which one obtains by optimizing the contraction sequence, are given in Table~\ref{tab:complexity}. Generally speaking, it is favorable to have a narrow causal cone, which explains why the classical costs for the 1D modified binary and ternary MERA are smaller than those of the plain binary MERA. On the other hand, for a given bond dimension $\chi$, the binary MERA can encode more entanglement than the other two 1D MERA types and generally achieves higher accuracy.

The costs shown in Table~\ref{tab:complexity} refer to one evaluation of the global energy gradient on a classical computer, or equivalently, a single update of all tensors in the Evenbly-Vidal algorithm \cite{Evenbly2009-79}. For a homogeneous MERA, this cost, like the number of different tensors, is linear in $T$. For heterogeneous MERA, the total number of tensors and the cost are proportional to $\sum_{\tau=1}^T b^\tau \sim N$. The table shows the classical costs for MERA with full tensors (fMERA) because, on classical computers, there is not much to gain by exploiting the Trotter structure of TMERA tensors unless one is willing to introduce approximations.

\subsection{Quantum computation time complexity}
\begin{table*}[t]
	\setlength{\tabcolsep}{1.57ex}
	\begin{tabular}{|l | c c | c c | c c c | c c|}
	 \hline
	 \multicolumn{1}{|c|}{MERA type}
           & \multicolumn{2}{c|}{Properties} & \multicolumn{2}{c|}{Number of qubits} & \multicolumn{3}{c|}{Times, homogeneous}       & \multicolumn{2}{c|}{Times, heterog.}\\
	                         & $b$ & $A$       & register & auxiliary & classical      & quantum          & f-quantum      & classical & quantum\\
	 \hline
	 1D binary               & 2 & 3           & $4q$     & $1$       & $\O(2^{9q}T)$  & $\O(q(tT)^2)$  & $\O(2^{8q}T^2)$  & $\O(2^{9q}N)$  & $\O(q(tT)^2N)$\\
	 1D mod.\ binary         & 2 & 2           & $3q$     & $2$      & $\O(2^{7q}T)$  & $\O(q(tT)^2)$  & $\O(2^{8q}T^2)$  & $\O(2^{7q}N)$  & $\O(q(tT)^2N)$\\
	 1D ternary              & 3 & 2           & $4q$     & $2$      & $\O(2^{8q}T)$  & $\O(q(tT)^2)$  & $\O(2^{8q}T^2)$  & $\O(2^{8q}N)$  & $\O(q(tT)^2N)$\\
	 2D $2\times 2\mapsto 1$ & 4 & $3\times 3$ & $14q$    & $2$      & $\O(2^{28q}T)$ & $\O(q(tT)^2)$  & $\O(2^{16q}T^2)$ & $\O(2^{28q}N)$ & $\O(q(tT)^2N)$\\
	 2D $3\times 3\mapsto 1$ & 9 & $2\times 2$ & $15q$    & $4$      & $\O(2^{16q}T)$ & $\O(q(tT)^2)$  & $\O(2^{20q}T^2)$ & $\O(2^{16q}N)$ & $\O(q(tT)^2N)$\\
	\hline
	\end{tabular}
	\caption{\label{tab:complexity}\textbf{Computational complexity.} For the five 1D and 2D MERA network structures shown in Fig.~\ref{fig:MERAs}, this table states the number of qubits needed for the variational quantum eigensolver and compares the time complexity for the quantum measurement of energy gradients to the evaluation in the classical algorithms.
	Columns 2 and 3 show the branching ratio $b$ and the cross section $A$ of the causal cone for local operators like the considered Hamiltonian interaction terms.
	Columns 4 and 5 show the number of qubits needed in the quantum processor register and the number of additional auxiliary qubits (per layer) needed to avoid circuit reprogramming in the gradient evaluations for homogeneous MERA.
	Columns 6--8 concern homogeneous MERA. Column 6 shows the classical computation times for fMERA, which agree with those for TMERA as, on a classical computer, the Trotter structure of tensors does not admit substantial gains. Columns 7 and 8 show the quantum computation times for TMERA and fMERA.
	Columns 9 and 10 show the classical and the quantum computation times for heterogeneous TMERA.}
\end{table*}
For the hybrid quantum-classical TMERA algorithm, the width of the causal cone is not as decisive for the computation costs. It determines primarily the number of qubits that need to be simultaneously in the interaction zone of the computer. Also, the specific network structure inside the cone, which influences the classical contraction costs, does not affect the scaling of the quantum computation costs.
As discussed in Sec.~\ref{sec:Trotterization}, the quantum costs for evaluating the TMERA expectation value of a local interaction term $\hh_i$ is proportional to $tT$.
For translation-invariant systems and homogeneous TMERA, the expectation value of the entire Hamiltonian $\hH=\sum_i\hh_i$ can in fact be measured in one go, i.e., with time $\O(tT)$. As there are $\N=\O(qtT)$ different Trotter gates, one needs $\O(q(tT)^2)$ time to measure the energy gradient.

For heterogeneous TMERA, there are $\N=\O(qtN)$ different Trotter gates with $\O(qtN/b^\tau)$ located in layer $\tau$. To measure the gradient with respect to one Trotter gate in layer $\tau$, we need to measure the $\O(b^\tau)$ expectation values for all local interactions terms that it affects. Hence, one needs $\sum_{\tau=1}^T\O(tT\, b^\tau\, qtN/b^\tau)=\O(q(tT)^2N)$ time for the gradient measurement of heterogeneous TMERA.

The comparison to the classical computation costs (Sec.~\ref{sec:ComplexityClassic}) is not trivial. In particular, one needs to take into account how many measurement samples are needed to reach a certain accuracy $\epsilon$. As described in Sec.~\ref{sec:CostAccuracy}, the comparison can be done by expressing the bond dimension $\chi$ as well as $qt^2$ in terms of $\epsilon$. For critical models, they are related by power laws. The comparison for a specific model is discussed in Appx.~\ref{sec:Complexity-BLBQ1}.

The classical component of the hybrid TMERA eigensolver controls the gradient evaluation and steers the gradient-based energy minimization. Its time complexity is always subleading to the quantum time complexity: For a TMERA with $\N$ different Trotter gates, the classical component operates on a vector space of dimension $\O(\N)$. The classical time complexity for one iteration of gradient descent or L-BFGS is just $\O(\N)$.

Although it is much less efficient than TMERA, one can also optimize fMERA using a quantum computer. A disentangler that acts on $n$ renormalized sites can be decomposed exactly into a circuit of $\O(2^{2nq})$ single-qubit and CNOT gates \cite{Barenco1995-52,Moettoenen2004-93,Shende2006-25}. An exact representation of an isometry that maps $n$ sites into $m>n$ requires $\O(2^{(n+m)q})$ single-qubit and CNOT gates \cite{Itan2016-93}. The time cost for the energy gradient measurement of a homogeneous fMERA is then obtained by squaring the largest number of gates per tensor and multiplying by $T^2$.

\subsection{Quantum space complexity with resets}
Concerning the quantum space complexity, we have to distinguish two settings. In the first, qubits corresponding to renormalized sites that leave the causal cone are reset to the reference state $|0\ket$ in order to reuse them for the implementation of isometries in the next layer transition $\tau\mapsto\tau-1$. The distribution of measurement results in the evaluation of energy gradients and local observables like $\bra\Psi_i|\hh_i|\Psi_i\ket$ is the same with and without such mid-circuit resets. Qubit rests can be implemented through a projective measurement followed by a subsequent $\pi$ rotation conditioned on the measurement result or through driven-dissipative reset schemes \cite{Reed2010-96,Magnard2018-121,Egger2018-10,Schindler2011-332,Gaebler2021-104}.

The number of qubits needed in the register is primarily determined by the cross section $A$ of the causal cone. More precisely, from the sequence of contractions inside each layer of the causal cone, one can determine how many renormalized sites (groups of $q$ qubits) are needed at any point in time. One can reduce this number by not insisting on parallel execution of gates, but shifting the MERA tensors of a layer-transition map temporally so that some qubits can already be reset before the application of further gates. The results for the considered MERA networks are shown in Table~\ref{tab:complexity}.

As discussed in Sec.~\ref{sec:TranslationInv}, a major computation time reduction for the homogeneous MERA is achieved by taking appropriate convex combinations of the layer-transition maps like $\hU_L$ and $\hU_R$ such that the energy density for the full (infinite) system is obtained in one go. This can be done either with classical randomness or, if one wants to avoid the corresponding reprogramming of pulse sequences in the experimental evaluation of gradients, by introducing auxiliary qubits. In the latter approach, the number of required auxiliary qubits per MERA layer is proportional to $\log_2$ of the number of transition maps. The auxiliary qubits are also amenable to qubit rests.

\subsection{Quantum space complexity without resets}
The experimental evaluation of observables and energy gradients can be made more time efficient using quantum amplitude estimation (QAE) \cite{Knill2007-75,Wang2019-122}. For the preparation $|\Psi_i\ket=\hU_i|0,\dotsc,0\ket$ of a causal-cone state, QAE requires application of powers $(\hU_i)^m$. In this approach, one cannot employ the resetting of qubits that leave the causal cone, because subsequent factors $\hU_i$ will again act on them. Without the resets, every layer transition $\tau\mapsto\tau-1$, requires $(b-1)Aq$ new qubits, initialized in the reference state $|0\ket$ in order to realize the isometries. For a TMERA with $T$ layers, one then needs a total of
\begin{equation}
	\sim (b-1)AqT
\end{equation}
qubits, i.e., a number that grows logarithmically in the total system size $N$. In ion-trap systems, one can use shuttling \cite{Rowe2002-2,Hensinger2006-88,Walther2012-109} to move currently used qubits in and out of the quantum register.

\subsection{Comparison of time complexities for critical spin-1 chains}\label{sec:Complexity-BLBQ1}
To directly compare the time complexities of the classical fMERA and the quantum-classical TMERA algorithms, one has to take into account the number of measurement samples needed in each optimization step of the TMERA algorithm. As discussed in Sec.~\ref{sec:CostAccuracy}, the required number of samples scales with the energy accuracy $\epsilon$. Employing QAE \cite{Knill2007-75,Wang2019-122}, the total time complexity per iteration is $\O(t^2 q \log(1/\epsilon) /\epsilon)$. The time complexity per iteration for a classical fMERA simulation is $\O(\chi^r)=\O(2^{rq})$, where the exponent $r$ depends on the MERA type as shown in Table~\ref{tab:complexity}.
\begin{figure}[t]
	\includegraphics[width=0.98\columnwidth]{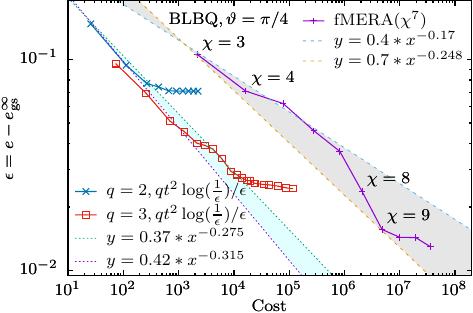}
	\caption{\label{fig:Complexity-BLBQ1}\textbf{Comparison of classical and VQE time complexities.} Double-logarithmic plot of the energy accuracy $\epsilon$ as a function of the total time complexity per iteration for the quantum-classical TMERA algorithm (lower-left legend) and the corresponding classical fMERA simulation (upper-right legend) applied to the critical bilinear-biquadratic spin-1 model \eqref{eq:H_blbq}. We use homogeneous modified binary MERA with $T=6$ layers. For the TMERA curves, the number of Trotter steps $t$ is gradually increased at fixed $q=2$ and $3$, respectively. For the fMERA curve, the bond dimension $\chi=2^q$ increases by one from point to point. In both cases, $\epsilon$ and the (optimal) cost are related by a power law as indicated by dashed lines. The data suggests a polynomial advantage of the TMERA algorithm.}
\end{figure}

Figure~\ref{fig:Complexity-BLBQ1} provides numerical results for  the bilinear-biquadratic spin-1 chain \cite{Uimin1970-12,Lai1974-15,Sutherland1975-12,Laeuchli2006-74,Binder2020-102}
\begin{equation}\label{eq:H_blbq}
 \hH = \sum_i\left[\cos\vartheta (\hat{\vec{S}}_i \cdot \hat{\vec{S}}_{i+1}) + \sin\vartheta (\hat{\vec{S}}_i \cdot \hat{\vec{S}}_{i+1})^2\right]
\end{equation}
at the critical Uimin-Lai-Sutherland point $\vartheta=\pi/4$. We choose this model because it corresponds to a conformal field theory with central charge $c=2$ \cite{Itoi1997-55} and, hence, features significant entanglement. The figure shows the energy accuracy $\epsilon$ as a function of the classical fMERA and quantum-classical TMERA computation cost per iteration. In both cases, the energy accuracy follows a power law with a moderate polynomial advantage for TMERA. For the TMERA curves with fixed $q$, $\epsilon$ decreases until reaching the accuracy of the fMERA with bond dimension $\chi=2^q$. In practical simulations, one should hence increase $q$ upon approaching the saturation, in order to follow the power-law decay as indicated by the dashed lines. The fMERA computation cost can be written as $\O(\chi^r)=\O(\epsilon^{-r/\beta})$, where the factor $1/\beta$ in the exponent captures the relation between bond dimension $\chi$ and accuracy $\epsilon$. As the exponent $r$ is very large ($r\geq 16$) for 2D MERA, the quantum-classical TMERA algorithm will substantially outperform the classical fMERA simulations for models in $\geq 2$ spatial dimensions.

\section{Layer-transition maps for homogeneous MERA}\label{sec:homogMERA}
\begin{figure*}[t]
	\includegraphics[width=0.99\textwidth]{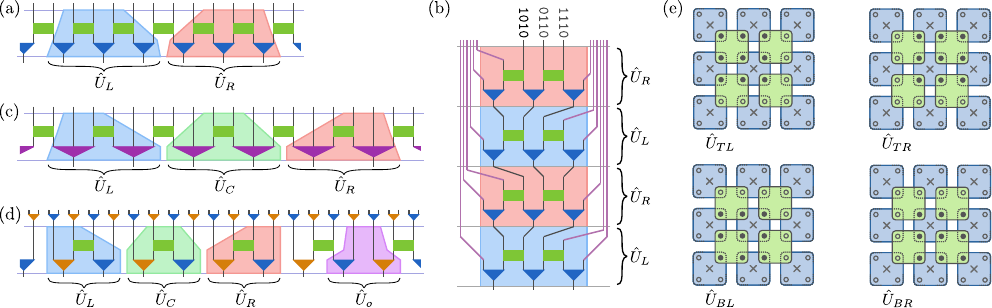}
	\caption{\label{fig:MERAtransitions}\textbf{Layer transition maps.} (a) The 1D binary MERA has causal-cone width $A=3$ and two layer-transition maps $\hU_L$ and $\hU_R$. Panel (b) gives an example for the sequence of transition maps in a particular causal cone, here, in correspondence with Fig.~\ref{fig:MERAs}a. For the binary MERA, the sequence can be determined from the binary representation of the index (four least significant digits shown) with $0$ corresponding to $\hU_L$ and $1$ to $\hU_R$. (c) The 1D ternary MERA has $A=2$ and three transition maps $\hU_L$, $\hU_C$, and $\hU_R$. (d) The 1D modified binary MERA has $A=2$ and one needs to distinguish even and odd bonds. Transition maps $\hU_L$ and $\hU_R$ map from even to odd bonds, $\hU_C$ from an even to an even bond, and $\hU_o$ from an odd to an even bond. (e) The 2D $2\times 2\mapsto 1$ MERA \cite{Cincio2008-100} has $A=3\times 3$ and four layer-transition maps $\hU_{TL}$, $\hU_{TR}$, $\hU_{BL}$, and $\hU_{BR}$. The renormalized causal-cone sites $\A_i^{(\tau)}$ that the maps act on are indicated by crosses, the sites that leave the causal cone during/after the map are indicated by empty circles, and the sites $\A_i^{(\tau-1)}$ that remain in the causal cone after the preparation step are indicated by filled circles. The layer-transition maps differ in the positions of the latter.}
\end{figure*}
Let us discuss in more detail, how classical sampling or auxiliary qubits can be employed to realize convex combinations of layer-transition maps in the preparation. In this way, the spatially averaged $A$-site density matrices of homogeneous TMERA can be prepared on the quantum computer in $\O(tT)$ time. Hence, the energy density or a component of the energy gradient can be obtained in $\O(tT)$ time.

\subsection{1D binary MERA}\label{sec:homogMERA1d}
This case was already discussed shortly in Sec.~\ref{sec:TranslationInv}. For the 1D binary MERA, local operators have causal cones of width $A=3$. The sequence of layer-transition maps $\hU_L$ and $\hU_R$ for the causal cone of sites
\begin{equation}\label{eq:siteBlock}
	\A_i:=\{i,\dotsc,i+A-1\}
\end{equation}
on the physical lattice can be deduced from the binary representation $i_1 i_2 i_3 \dots$ of $i$ with $i_\tau\in\{0,1\}$, where $i_1$ is the least significant bit. For the choice shown in Figs.~\ref{fig:MERAs}a and \ref{fig:MERAtransitions}a, $i_\tau=0$ or $1$ means that we progress from layer $\tau$ to layer $\tau-1$ by applying transition map $\hU_L\equiv \hU_0$  or $\hU_R\equiv\hU_1$, respectively. Specifically, the causal-cone state for sites $\A_i$ reads
\begin{equation}\label{eq:transitionPsi}
	|\Psi_i\ket = \hU_{i_1}\dotsb\hU_{i_T}\big(|T\ket^{\otimes A}\big),
\end{equation}
where state $|T\ket$ for a single renormalized site defines the so-called top tensor of the MERA. Here and in the following, we do not explicitly denote the $\tau$ dependence of the layer-transition maps and use the convention that operators act on the $A$ \emph{active} sites in the causal cone and the remaining \emph{inactive} sites are left untouched as indicated in Fig.~\ref{fig:MERAtransitions}b.

For quantities like the energy density, we wish to evaluate the spatial average $\dm=\frac{1}{N}\sum_i\dm_i$ of the reduced density matrices $\dm_i:=\Tr_{\A_i^\bot}|\Psi_i\ket\bra\Psi_i|$ for the blocks of sites $\A_i$. The partial trace is over the inactive sites (outside the causal cone) and will not be denoted explicitly in the following. Starting from $\dm^{(T)}:=(|T\ket\bra T|)^{\otimes A}$, we can recursively define
\begin{subequations}\label{eq:transitionChannel}
\begin{equation}\label{eq:transitionChannel-step}
	\dm^{(\tau-1)}=\frac{1}{2}\sum_{i_\tau}\hU_{i_\tau} \dm^{(\tau)} \hU_{i_\tau}^\dag
\end{equation}
such that
\begin{equation}
	\dm^{(0)}=\frac{1}{2^T}\sum_{i_1,\dotsc,i_T}\hU_{i_1}\dotsb\hU_{i_T}\dm^{(T)}\hU_{i_T}^\dag\dotsb\hU_{i_1}^\dag=\dm.
\end{equation}
\end{subequations}
$\hU_L$ and $\hU_R$ are related to each other by a site permutation.

Hence, the quantum channel \eqref{eq:transitionChannel-step} can be implemented through a Stinespring dilation \cite{Stinespring1955-6,Nielsen2000} with a single auxiliary qubit per layer. The auxiliary qubit is initialized in the state $(|0\ket+|1\ket)/\sqrt{2}$ and used for a controlled site permutation such that, after tracing out the auxiliary qubit, one obtains the channel \eqref{eq:transitionChannel-step}.

Alternatively, one can use classical random numbers. Selecting in each layer transition $\hU_L$ or $\hU_R$ with equal probability, also implements the channel \eqref{eq:transitionChannel-step}. Mathematically, this scheme is completely equivalent to the approach using auxiliary qubits and results in the same distribution of measurement results in the evaluation of observables and energy gradients: Whether one prepares states $|\Psi_1\ket$ and $|\Psi_2\ket$ with 50:50 probability or prepares
$\left(|\Psi_1\ket\otimes |0\ket + |\Psi_2\ket \otimes |1\ket\right)/\sqrt{2}$ with an auxiliary qubit, the probability to observe outcome $i$ associated with projector $\hat{P}_i$ is in both cases $\big(\bra\Psi_1|\hat{P}_i|\Psi_1\ket + \bra\Psi_2|\hat{P}_i|\Psi_2\ket\big)/2$.
Experimentally, the measurements need to be repeated until reaching a required accuracy $\epsilon$. Practically, the approach using classical sampling involves a reprogramming of the experimental pulse generator hardware for each new sequence of transition maps. If this reprogramming is slow, the scheme with auxiliary qubits may be preferable as it employs the same quantum circuit in every measurement iteration.

\subsection{1D ternary MERA}
We can proceed along the same lines for the 1D ternary MERA. The only modifications are due to, now, having the three transition maps $\hU_L\equiv \hU_0$, $\hU_C\equiv\hU_1$, and $\hU_R\equiv\hU_2$ shown in Fig.~\ref{fig:MERAtransitions}c. Correspondingly, we now use the ternary representation $i_1 i_2 i_3 \dots$ of the site index $i$ with $i_\tau\in\{0,1,2\}$ and replace the prefactors $1/2$ in Eqs.~\eqref{eq:transitionChannel} by $1/3$. For the Stinespring dilation, we now require two auxiliary qubits per layer, which are, for example, initialized in the state $(|0,0\ket+|1,0\ket+|0,1\ket)/\sqrt{3}$ to perform the controlled site permutations that relate the three unitaries $\hU_{i_\tau}$.

\subsection{2D \texorpdfstring{$2\times 2\mapsto 1$}{2x2-to-1} MERA}
Also the 2D $2\times 2\mapsto 1$ MERA can be treated similarly. Now, we use binary representations $x_1 x_2 x_3 \dots$ and $y_1 y_2 y_3 \dots$ for the $x$ and $y$ coordinates of site $i$ on the square lattice with $x_\tau,y_\tau\in\{0,1\}$. There are four layer-transition maps $\hU_{TL}\equiv \hU_{0,0}$, $\hU_{TR}\equiv \hU_{1,0}$, $\hU_{BL}\equiv \hU_{0,1}$, $\hU_{BR}\equiv \hU_{1,1}$, where $\hU_{TL}$ corresponds to the causal cone continuing along the inner (top-left corner) $3\times 3$ square indicated in Fig.~\ref{fig:MERAtransitions}e. The layer-transition channel reads
\begin{equation}\label{eq:transitionChannel-step2x2}
	\dm^{(\tau-1)}=\frac{1}{4}\sum_{x_\tau,y_\tau}\hU_{x_\tau,y_\tau} \dm^{(\tau)} \hU_{x_\tau,y_\tau}^\dag.
\end{equation}
Its Stinespring dilation can be implemented with two auxiliary qubits, each initialized in $(|0\ket+|1\ket)/\sqrt{2}$, to realize the controlled site permutations that relate the maps $\hU_{x_\tau,y_\tau}$ to each other. The first auxiliary qubit can be used to permute rows of sites and the second to permute columns of sites on the inner $4\times 4$ square in Fig.~\ref{fig:MERAtransitions}e.

\subsection{1D modified binary MERA}
The cases of the 1D modified binary MERA and the 2D $3\times 3\mapsto 1$ MERA are bit more involved.

First, note that the layer-transition channel \eqref{eq:transitionChannel-step} for the (unmodified) 1D binary MERA can also be explained as follows: Let $\A^{(\tau)}_i$ denote the block of renormalized sites $\{i,\dotsc,i+A-1\}$ after $\tau$ renormalization steps, i.e., at the interface of layers $\tau$ and $\tau+1$; $\A^{(0)}_i$ agrees with the block $\A_i$ of physical sites in Eq.~\eqref{eq:siteBlock}. Let $\dm^{(\tau)}_i$ denote the reduced density matrix for $\A^{(\tau)}_i$, constructed from all MERA tensors in the causal cone of $\A^{(\tau)}_i$ which ranges from layer $\tau+1$ to the final layer $T$. 
Given all density matrices $\dm^{(\tau)}_i$ for layer $\tau$, we want to compute those for layer $\tau-1$. The density matrices $\dm^{(\tau-1)}_j$ and $\dm^{(\tau-1)}_{j+1}$ for the two blocks $\A^{(\tau-1)}_j$ and $\A^{(\tau-1)}_{j+1}$ whose causal cones contain $\A^{(\tau)}_i$, are obtained by conjugating $\dm^{(\tau)}_i$ with the appropriate layer-$\tau$ transition maps
\begin{equation}
	\dm^{(\tau-1)}_j=\hU_L \dm^{(\tau)}_i \hU_L^\dag \quad\text{and}\quad
	\dm^{(\tau-1)}_{j+1}=\hU_R \dm^{(\tau)}_i \hU_R^\dag.
\end{equation}
Now, the spatially averaged density matrix $\dm^{(\tau-1)}=\frac{b^{\tau-1}}{N}\sum_j\dm^{(\tau-1)}_j$ is obtained by averaging over all sites $i$ in layer $\tau$ and the two corresponding sites in layer $\tau-1$. Thus, we recover Eq.~\eqref{eq:transitionChannel-step}.

The 1D modified binary MERA with $A=2$ can be addressed similarly. In this case, not all bonds are equivalent. Depending on whether we are on an even or an odd bond, we apply either one of the three layer-transition maps $\hU_L,\hU_C,\hU_R$ or the map $\hU_o$, respectively, as shown in Fig.~\ref{fig:MERAtransitions}d. While $\hU_L$ and $\hU_R$ move the causal cone from an even to an odd bond, $\hU_C$ maps from an even to an even bond, and $\hU_o$ maps from an odd to an even bond. Let $\dm^{(\tau)}_e$ denote the bond density matrix averaged over all even bonds for layer $\tau$ and $\dm^{(\tau)}_o$ the average over all odd bonds. Then, the same considerations as above lead to the layer transitions
\begin{subequations}
\begin{align}
	\dm_e^{(\tau-1)} &= \frac{1}{2}\left( \hU_C \dm_e^{(\tau)} \hU_C^\dag + \hU_o \dm_o^{(\tau)} \hU_o^\dag \right),\\
	\dm_o^{(\tau-1)} &= \frac{1}{2}\left( \hU_L \dm_e^{(\tau)} \hU_L^\dag + \hU_R \dm_e^{(\tau)} \hU_R^\dag \right).
\end{align}
\end{subequations}
For the evaluation of energy densities and gradients we want to prepare the spatially averaged density matrices $\dm_e^{(0)}$ and $\dm_o^{(0)}$ for even and odd bonds of the physical lattice.

To realize this in the approach using auxiliary qubits, we can first introduce an additional flag qubit in the register of the quantum computer which indicates whether we are on an even or odd bond, i.e., progressing in the preparation direction, we want to prepare the states
\begin{equation}\label{eq:1DmodBin-withFlag}
	\dm^{(\tau)}:=\frac{1}{2}\left(|0\ket\bra 0|_f\otimes \dm_e^{(\tau)} + |1\ket\bra 1|_f\otimes \dm_o^{(\tau)}\right).
\end{equation}
For these, the layer-transition channel reads
\begin{gather}
	\dm^{(\tau-1)}=\frac{1}{2}\sum_{k=1}^4 \hL_k \dm^{(\tau)} \hL_k^\dag\quad\text{with}\\\nonumber
	\hL_1=|0\ket\bra 0|_f\otimes \hU_C,\quad \hL_2=|0\ket\bra 1|_f\otimes \hU_o\\\nonumber
	\hL_3=|1\ket\bra 0|_f\otimes \hU_L,\quad \hL_4=|1\ket\bra 0|_f\otimes \hU_R.
\end{gather}
It can be implemented by a Stinespring dilation that employs two auxiliary qubits per layer as specified in the second row of Table~\ref{tab:complexity}.

\section{Gradient evaluation and Riemannian optimization}\label{sec:Gradients}
The goal of the variational quantum eigensolver is to minimize the energy expectation value $E=\bra\Psi|\hH|\Psi\ket$ over a TMERA variety $\{|\Psi\ket\}$, where the TMERA is characterized by the network structure, bond dimensions, and the tensor Trotterization as previously discussed. This minimization can be carried out by evaluating energy gradients and employing them in gradient descent methods or, preferably, quasi-Newton methods like the limited-memory Broyden–Fletcher–Goldfarb–Shanno (L-BFGS) algorithm \cite{Nocedal2006,Liu1989-45}.

\subsection{Gradients in the CNOT and CAN parametrizations}
In the CNOT and CAN parametrizations, the Trotter gates are expressed in terms of single and two-qubit rotations $\hR_\hs(\theta)=e^{-\mri\theta \hs/2}$ [Eq.~\eqref{eq:rotation}].
The rotation angles $\theta$ parametrize the TMERA variety. To compute the energy derivative for one of these angles, we can write the energy expectation value in the form
\begin{equation}\label{eq:Etheta}
	E(\theta)=\Tr\left(\hA [\hR_\hs^\dag(\theta)\otimes \id_\bot]\hB[\hR_\hs(\theta)\otimes \id_\bot]\right),
\end{equation}
where the Hermitian operators $\hA$ and $\hB$ comprise the remaining tensors of $\bra\Psi|$, $|\Psi\ket$, and the Hamiltonian. For brevity of notation, in the following, we will drop the ``$\otimes \id_\bot$'', which indicates that $\hR_\hs(\theta)$ only acts on a subspace corresponding to one or two qubits. The derivative is
\begin{equation}\label{eq:Etheta_D}\textstyle
	\partial_\theta E(\theta)=\frac{\mri}{2}\Tr\left(\hA \hR_\hs^\dag(\theta) [\hs,\hB]\hR_\hs(\theta)\right).
\end{equation}
For the Hermitian and unitary operators $\hs$, $\hR_\hs(\pm\frac{\pi}{2})=(\id\mp \mri\hs)/\sqrt{2}$ and, hence,
\begin{equation}\label{eq:commute_sigma}\textstyle
	\mri[\hs,\hB]=\hR_\hs^\dag(\frac{\pi}{2})\hB\hR_\hs(\frac{\pi}{2}) -\hR_\hs^\dag(-\frac{\pi}{2})\hB\hR_\hs(-\frac{\pi}{2}).
\end{equation}
such that Eq.~\eqref{eq:grad} follows. In this way, the gradient can be evaluated on the quantum computer by measuring energy expectation values \cite{Li2017-118,Guerreschi2017_01,Mitarai2018-98}.

For homogeneous TMERA, the same rotation occurs multiple times as tensors are repeated in the translation-invariant MERA layers. Applying the product rule, this just means that the derivative $\partial_\theta E$ will contain one term for each occurrence of the rotation $\hR(\theta)$. This sum can be evaluated efficiently as described in the main text and Appx.~\ref{sec:homogMERA}.

With the gradient in hand, one can apply standard implementations of gradient descent or quasi-Newton methods like L-BFGS.
\begin{figure*}[t]
\hrule
\vspace{0.5em}
\begin{algorithmic}[1]
\State $\vu_0\in\M$, \ $\veps>0$, \ $\ell\in\mathbb{N}$, \ $0<c_1<\frac{1}{2}<c_2<1$, \ $\gamma=1$, \ $k=0$, \ $m=0$  \Comment{Initialization.}
\While{$\|\vg_k\|>\veps$} \Comment{Riemannian gradient $\vg_k$ at $\vu_k$ determined according to Eq.~\eqref{eq:Riem_grad}.}
    \State $\vp_k = -\vg_k$ \Comment{Lines 3-11 determine the search direction $\vp_k=-\tilde{H}_k\vg_k$ with Eq.~\eqref{eq:BFGS-update}.}
    \For{$i=k-1,k-2,\dotsc,m$} \Comment{Compute $-V^\dag_{m}V^\dag_{m+1}\dotsb V^\dag_{k-1}\vg_k$ and $\xi_i:=-(\vs_i,V^\dag_{i}\dotsb V^\dag_{k-1}\vg_k)$.}
      \State $\xi_i \gets \rho_i(\vs_i,\vp_k)$
      \State $\vp_k \gets \vp_k - \xi_i \vy_i$
    \EndFor
    \State $\vp_k \gets \gamma\vp_k$ \Comment{Multiply with $\tilde{H}_{m-1}:=\gamma\id$.}
    \For{$i=m,m+1,\dotsc,k-1$} \Comment{Multiply with $V_{k-1} \dotsb V_{m}$ and insert $\rho_i\vs_i \vs_i^\dag$ terms.}
      \State $\vp_k \gets \vp_k -\rho_i\vs_i (\vy_i,\vp_k) + \vs_i \xi_i$
    \EndFor \Comment{Now, $\vp_k=-\tilde{H}_k\vg_k$.}
    \State Do line search to find $\tau_k\in\RR$ that satisfies the Wolfe conditions. With $\vu_{k+1}:=\vr_{\vu_{k},\vp_k}(\tau_k)$:
    \State \hspace{4ex} $E( \vr_{\vu_{k},\vp_k}(\tau_k) )\leq E(\vu_k) + c_1 \tau_k (\vg_k,\vp_k)$, 
    \State \hspace{4ex} $\partial_\tau E(\vr_{\vu_{k},\vp_k}(\tau))|_{\tau=\tau_k} \geq c_2 \partial_\tau E(\vr_{\vu_{k},\vp_k}(\tau))|_{\tau=0}$.\label{line:Wolfe2}
    \State $\vu_{k+1} = \vr_{\vu_{k},\vp_k}(\tau_k)$ \Comment{For the following, $T_k:=T_{\vu_k,\vp_k}(\tau_k)$ from Eq.~\eqref{eq:Riem_transport}.}
    \State $\vs_k=T_k\tau_k\vp_k$, \ $\vy_k=\vg_{k+1}-T_k\vg_k$, \
           $\rho_k=1/(\vs_k,\vy_k)$, \ $\gamma \gets (\vs_k,\vy_k)/\|\vy_k\|^2$
    \State $m \gets \max(k-\ell,0)$
    \State Discard $(\rho_{m-1},\vs_{m-1},\vy_{m-1})$ from memory if $m>0$.
    \For{$i=m,m+1,\dotsc,k-1$} \Comment{Transport $\{\vs_i\}$ and $\{\vy_i\}$ from $\mc{T}_{\vu_k}$ to $\mc{T}_{\vu_{k+1}}$.}
      \State $\vs_i \gets T_k\vs_i$, \ $\vy_i \gets T_k\vy_i$
    \EndFor
    \State $k\gets k+1$
\EndWhile 
\end{algorithmic}
\hrule
\caption{\label{alg:L-BFGS}Riemannian version of the L-BFGS algorithm to minimize (T)MERA energies, adapted from Ref.~\cite{Huang2015-25}. The optimization is started at a point $\vu_0\in\M$ and stopped when the gradient norm falls below $\veps$. For the approximation $\tilde{H}_k$ of the inverse Hessian, $\ell$ vector pairs $(\vs_i,\vy_i)$ with $i=k-\ell,\dotsc,k-1$ are kept in memory. The constants $c_1$ and $c_2$ enter the Wolfe conditions.}
\end{figure*}

\subsection{An alternative Riemannian version of the optimization}\label{sec:Riemannian}
Instead of employing an explicit parametrization of the Trotter gates, one can formulate the problem as a minimization over the manifold
\begin{equation}\label{eq:Riem_manifold}
	\M:=\U(4)^{\times \N}
\end{equation}
formed by the product of the unitary groups for the $\N$ Trotter gates of the TMERA. For homogeneous TMERA, repeated gates are counted once. This Riemannian approach turns out to have somewhat improved convergence properties as discussed in Appx.~\ref{sec:localMinimia}.
Note that one can take unitary gauge freedoms on the inputs of the Trotter gates into account and consider quotient groups $\U(4)/(\U(2)\times \U(2))$. For the simplicity of notation, we stick to the full $\U(4)$ in the following.

For the optimization, we can regard $\M$ as embedded in the Euclidean space
\begin{equation}\label{eq:Riem_Embedd}
	\E=\operatorname{End}(\CC^4)^{\times\N}\simeq\RR^{32\N},
\end{equation}
which is the space of the Trotter gates without the unitarity constraint. Let $\vu\in\M\subset\E$ denote the vector that contains the matrix elements of all gates and $E(\vu)=\bra\Psi(\vu)|\hH|\Psi(\vu)\ket$ the energy functional. To apply gradient-based optimization algorithms in this setting, we need to compute the derivative $\partial_\vu E(\vu)$, project it onto the tangent space $\mc{T}_\vu$ of $\M$ at $\vu$ to obtain the gradient direction, construct retractions for line search, and vector transport to be able to sum gradient vectors from different points on the manifold. This is the program of Riemannian optimization as discussed generally in Refs.~\cite{Smith1994-3,Huang2015-25} and recently demonstrated for MERA in Refs.~\cite{Hauru2021-10,Luchnikov2021-23}.

In the following, consider a single unitary $\hu\in\U(n)$, with $n=4$ for the considered Trotter gates, and we employ the Euclidean metric (real part of the Hilbert-Schmidt inner product)
\begin{equation}
	(\hu,\hu'):=\Re\Tr(\hu^\dag\hu').
\end{equation}
The extension to the product manifold $\M$ is straightforward. As in Eq.~\eqref{eq:Etheta}, let us write the energy expectation value in the form
\begin{align}\nonumber
	E(\hu)&=\bra\Psi(\hu)|\hH|\Psi(\hu)\ket\\ \label{eq:Riem_Eu}
	      &=\Tr(\hA [\hu^\dag\otimes\id_\bot]\hB[\hu\otimes\id_\bot]),
\end{align}
where ``$\otimes \id_\bot$'' indicates that $\hu$ only acts on an $n$-dimensional subspace.
Then, the energy gradient in the embedding space $\operatorname{End}(\CC^{n})$ is
\begin{equation}
	\hd=2\Tr_\bot(\hB[\hu\otimes\id_\bot]\hA),
\end{equation}
where $\Tr_\bot$ is the partial trace over the subspace that $\hu$ does \emph{not} act on.
The gradient $\hd$ fulfills $\partial_\veps E(\hu+\veps\hw)|_{\veps=0}=(\hd,\hw)$ for all $\hw$. An element $\hw$ of the tangent space $\mc{T}_\hu$ for $\U(n)$ at $\hu$ needs to obey $(\hu+\veps \hw)^\dag(\hu+\veps \hw)=\id+\O(\veps^2)$, i.e., $\hu^\dag\hw+\hw^\dag\hu=0$. So, $\hu^\dag\hw$ needs to be skew-Hermitian and, hence,
\begin{equation}
	\mc{T}_\hu=\{\mri\hu\heta\,|\,\heta=\heta^\dag\in\operatorname{End}(\CC^{n})\}.
\end{equation}
The Riemannian energy gradient $\hg$ for the manifold $\U(n)$ at $\hu$ is obtained by projecting $\hd$ onto the tangent space such that $(\hw,\hg)=(\hw,\hd)$ for all $\hw\in\mc{T}_\hu$. This gives
\begin{equation}\label{eq:Riem_grad}
	\hg = (\hd - \hu\hd^\dag\hu)/2 \ \in \  \mc{T}_\hu.
\end{equation}
For a line search on the manifold, we need a retraction, i.e., a curve $\hr_{\hu,\hp}(\tau)$ on the manifold that starts from $\hu=\hr_{\hu,\hp}(0)$ in direction $\hp=\partial_\tau\hr_{\hu,\hp}(\tau)|_{\tau=0}\in \mc{T}_\hu$. We use
\begin{equation}\label{eq:Riem_retract}
	\hr_{\hu,\hp}(\tau):=e^{\tau \hp\hu^\dag}\hu\ \in\ \U(n).
\end{equation}
For quasi-Newton methods, we also need to compute differences of Riemannian gradients from different points on the manifold, specifically, for two points on a retraction. This is accomplished by vector transport, i.e., a map between the two corresponding tangent spaces. We use,
\begin{equation}\label{eq:Riem_transport}
	\hT_{\hu,\hp}(\tau)\hw:=e^{\tau\hp\hu^\dag}\hw \quad \text{for} \quad \hw\in\mc{T}_\hu,\ \tau\in\RR
\end{equation}
This gives an element of the tangent space at $\hr_{\hu,\hp}(\tau)$ and $\hT_{\hu,0}\hw=\hw$. The vector transport is isometric in the sense that $\big(\hT_{\hu,\hp}(\tau)\hw,\hT_{\hu,\hp}(\tau)\hw'\big)=(\hw,\hw')$ for all $\hp,\hw,\hw'\in\mc{T}_\hu$ and $\tau\in\RR$.

In analogy to Eq.~\eqref{eq:grad}, the Riemannian gradient \eqref{eq:Riem_grad} can be obtained on the quantum computer by measuring energies of TMERA where one Trotter gate is modified: The tangent space $\mc{T}_\hu$ is a real $n^2$-dimensional vector space, and we can choose a basis $\{\mri\hu\hs_j\,|\,j=1,\dots,n^2\}$ with Hermitian unitaries $\hs_j$, i.e., $\hs_j=\hs_j^\dag$ and $\hs_j^2=\id$.
With $(\hs_j,\hs_k)=\delta_{j,k}n$, we can expand the energy gradient in the form $\hg=\mri\sum_{j=1}^{n^2}\alpha_j\hu\hs_j/n$ and evaluate the expansion coefficients $\alpha_j$ using the energy expectation values \eqref{eq:Riem_Eu},
\begin{align}\nonumber
	\alpha_j&=(\hg,\mri\hu\hs_j)=(\hd,\mri\hu\hs_j)=\Re\Tr(\mri\hd^\dag\hu\hs_j)\\\nonumber
	&=-\mri\Tr\left(\hA\big[\hs_j\otimes \id_\bot,(\hu^\dag\otimes \id_\bot) \hB(\hu\otimes \id_\bot)\big]\right)\\
	&\textstyle= E\left(\hu\hR_{\hs_j}(-\frac{\pi}{2})\right) - E\left(\hu\hR_{\hs_j}(\frac{\pi}{2})\right).
\end{align}
For the third line, we have used Eq.~\eqref{eq:commute_sigma} and $\hR_\hs(\theta)=e^{-\mri\theta\hs/2}$.

To minimize TMERA energies, we can employ a Riemannian version of the L-BFGS algorithm. For the following, we return to the global optimization problem on the product manifold \eqref{eq:Riem_manifold} with the embedding space \eqref{eq:Riem_Embedd}, vectors $\vu\in\M$ comprising the matrix elements of all Trotter gates, and the Euclidean metric $(\vu,\vu')=\Re(\vu^\dag\vu')$. Similarly, gradients $\hg$, retractions $\hr$ etc.\ are now written in a vectorized form. In the Newton method, one generates a sequence of points $\vu_1,\vu_2,\dots\in\M$ that converges quadratically fast to a minimum of $E(\vu)$. In each step, one obtains a second-order model of $E(\vu)$ using the Riemannian gradient $\vg_k$ and the inverse Hessian $H_k$ at $\vu_k$. The vector $\vec{p}_k:=-H_k\vg_k\in \mc{T}_{\vu_{k}}$ that points from $\vu_k$ to the minimum of the quadratic model is used for an inexact line search, and one chooses the next point $\vu_{k+1}=\vr_{\vu_{k},\vp_k}(\tau_k)$ on the corresponding line (retraction curve) with $\tau_k\in\RR$ such that the Wolfe conditions are obeyed. The latter require that both the function value and the gradient norm decrease sufficiently, where, in the Riemannian version, one rather considers the energy derivatives in the search direction. The BFGS algorithm \cite{Nocedal2006,Huang2015-25} modifies this procedure, avoiding the costly evaluation of the Hessian. Instead, one updates a positive definite approximation $\tilde{H}_k\in \operatorname{End}(\mc{T}_{\vu_k})$ of the inverse Hessian. $\tilde{H}_{k+1}$ is determined by requiring that the gradient of the new quadratic model at $\vu_{k+1}$, evaluated at $\vu_k$, should agree with the actual $\vg_k$. This is equivalent to the secant equation $\vs_k=\tilde{H}_{k+1}\vy_k$ with $\vs_k:=T_k\tau_k\vp_k$ and the gradient change $\vy_k:=\vg_{k+1}-T_k\vg_k$, which are both elements of the tangent space $\mc{T}_{\vu_{k+1}}$ as $T_k:=T_{\vu_k,\vp_k}(\tau_k)$ denotes the vector transport. From the solution space of the secant equation, one chooses the matrix $\tilde{H}_{k+1}$ that is closest to $\tilde{H}_{k}$ in a suitable metric. Specifically, the BFGS update reads
\begin{equation}\label{eq:BFGS-update}
	\tilde{H}_{k+1} = V_k \tilde{H}_k V_k^\dag +\rho_k\vs_k \vs_k^\dag
\end{equation}
with $V_k:=(\id-\rho_k\vs_k\vy_k^\dag)T_k$ and $\rho_k:=1/(\vy_k,\vs_k)$.
Finally, the L-BFGS algorithm \cite{Nocedal2006,Huang2015-25} avoids the increasing cost of operating with $\tilde{H}_{k}$ by keeping the $\ell$ most recent triples $(\rho_k,\vs_k,\vy_k)$ in memory and computing an approximation of the inverse Hessian from them in each iteration.

The Riemannian L-BFGS algorithm that we employ, adapted from Ref.~\cite{Huang2015-25}, is shown in Fig.~\ref{alg:L-BFGS}. Note that, for numerical stability, after every retraction \eqref{eq:Riem_retract}, it may be necessary to project the resulting point onto the manifold in order to avoid the accumulation of small numerical errors. For example, this can be done by a singular value decomposition for every Trotter gate and setting all singular values to one. If this is done, the finite numerical precision for the vector transport \eqref{eq:Riem_transport} should have negligible effects, i.e., need not be corrected.

The (classical) computation costs for each iteration of the Riemannian L-BFGS algorithm are linear in the total number $\N$ of Trotter gates and linear in the number $\ell$ of retained terms (rank of $\tilde{H}_{k}$). The latter does not need to be scaled with the problem size, and we choose $\ell=9$ in all computations. Retractions \eqref{eq:Riem_retract}, vector transport \eqref{eq:Riem_transport} etc.\ can be applied separately for every Trotter gate, each corresponding to $2 n^2$ entries of the full $\vu$ vector with $n=4$. The cost is hence indeed $\O(n^3\ell \N)=\O(\N)$.

\section{Different Trotter-gate parametrizations and \texorpdfstring{$XX$}{XX}-TMERA}\label{sec:localMinimia}
\begin{figure}[t]
	\includegraphics[width=\columnwidth]{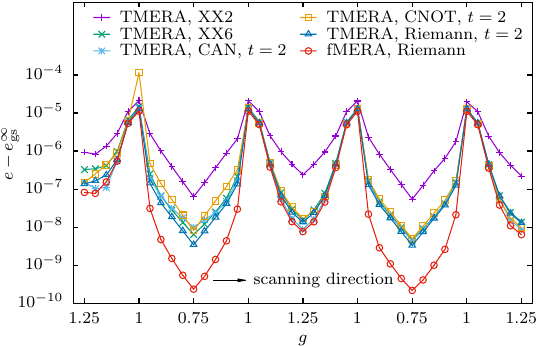}
	\caption{\label{fig:scanning}\textbf{Optimization and scanning with different parametrizations.} Starting from a product state at $g=1.25$, we scan forth and back on the interval $0.75\leq g\leq 1.25$, minimizing the MERA energy density $e$ for the 1D transverse-field Ising model with the L-BFGS algorithm. This is done for homogeneous modified binary MERA with $T=6$ layers and bond dimension $\chi=2^q=8$. The TMERA tensors consist of $t=2$ Trotter steps, and $XX$-TMERA are simulated once with two and once with six $XX$ Trotter steps per tensors.}
\end{figure}
The CNOT and CAN parametrizations for the Trotter gates of the TMERA are equivalent to the parametrization-free representation as $\U(4)$ unitaries, and, up to an irrelevant phase factor, one can transform between the parametrizations as described in Refs.~\cite{Shende2004-69,Kraus2001-63}. This equivalence is tested and confirmed by optimizing TMERA for the 1D transverse-field Ising model \eqref{eq:tIsing} while scanning forth and back on the parameter interval $g\in [0.75,1.25]$. Starting at $g=1.25$, the optimization was initialized by a product state with
	$|\phi\ket=\hR_{\hs^z}(\frac{\pi}{4})\hR_{\hs^y}(\frac{\pi}{4})\hR_{\hs^z}(\frac{\pi}{4})|0\ket$
on every site. Figure~\ref{fig:scanning} shows the accuracies of the energy densities $e$ during the scanning procedure for TMERA in the CNOT and CAN parametrizations as well as the parametrization-free form (``Riemannian''). The results are compared to the corresponding fMERA optimization.
The Euclidean L-BFGS algorithm \cite{Nocedal2006,Liu1989-45} was employed for the TMERA in the CAN and CNOT parametrizations and the Riemannian L-BFGS algorithm, as discussed in Appx.~\ref{sec:Riemannian}, was employed for the parametrization-free TMERA and fMERA. In all cases, the L-BFGS parameters were chosen as $\veps=10^{-12}$, $\ell=9$, $c_1=0.1$, and $c_2=0.9$.
\begin{figure}[t]
	\includegraphics[width=\columnwidth]{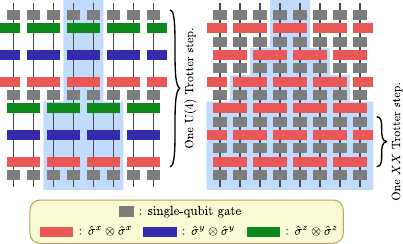}
	\caption{\label{fig:XX-TMERA}\textbf{$XX$-Trotterization.} Left: In the TMERA considered so far, tensors are Trotterized into regular circuits of general two-qubit Trotter gates $\in\U(4)$. For 1D systems, one Trotter step consists of such gates applied on all odd bonds and then on all even bonds (or vice versa). In the CAN parametrization, each $\U(4)$ Trotter gate is realized by a sequence of two single-qubit gates, followed by three Ising rotations generated by $\hs^\alpha\otimes\hs^\alpha$, and two final single-qubit gates. In every Trotter step, causal cones widen by at most four sites. Right: In the $XX$-TMERA, layers of single-qubit gates alternate with layers of $\hs^x\otimes\hs^x$ Ising gates on even and odd bonds, respectively. One $XX$ Trotter step contains two layers of Ising gates. In terms of the number of Ising gates, one $\U(4)$ Trotter step corresponds to three $XX$ Trotter steps. For the latter, causal cones grow correspondingly faster.}
\end{figure}

Although different TMERA parametrizations show differing energies in the early stages of the scanning procedure, they quickly converge. Without scanning, the parametrization-free TMERA is somewhat favorable. The explicit parametrizations in terms of rotation angles are more prone to getting stuck in local minima, whereas the local minima and saddle points for the parametrization-free form are entirely due to the structure of the TMERA manifold and the Hamiltonian. For random initial states, the parametrization-free form shows better convergence than the CNOT and CAN forms.
As a simple example, note that the product state $|0\ket^{\otimes N}$ is a stationary point in the energy landscape of the transverse Ising model \eqref{eq:tIsing} in the CAN parametrization.
The data in Fig.~\ref{fig:tIsing} was obtained by Riemannian optimization.

For the 1D TMERA, we chose the Trotter circuit of each tensor to consist of steps with $\U(4)$ gates on all odd qubit bonds and all even qubit bonds, alternatingly, as shown in Fig.~\ref{fig:TMERA}c and in the left panel of Fig.~\ref{fig:XX-TMERA}. The goal of this choice is to admit the generation of entanglement between any pair of qubits in a few Trotter steps. However, in the CNOT and CAN parametrizations, each Trotter gate features three elementary two-qubit gates; CNOT and $\hs^\alpha\otimes\hs^\alpha$ Ising rotations with $\alpha=x,y,z$, respectively. We have explored a different Trotterization approach, where layers of generic single-qubit gates alternate with $XX$ Ising rotations. One $XX$ Trotter step contains two layers of $XX$ Ising rotations on odd and even bonds, respectively. For the same computational cost, characterized by the total number of two-qubit Ising rotations, causal cones in the $XX$ Trotterization of the MERA tensors grow substantially faster than in the $\U(4)$ Trotterization. However, the benchmark simulations in Fig.~\ref{fig:scanning} show no enhanced approximation accuracy. The energies for six $XX$ Trotter steps per MERA tensor converge to approximately the same values as the energies for two $\U(4)$ Trotter steps per tensor. Improvements along these lines are a topic for future work.

\bibliographystyle{prsty.tb.title}

\end{document}